\documentclass[conference]{IEEEtran}
\IEEEoverridecommandlockouts
\usepackage{cite}
\usepackage{amsmath,amssymb,amsfonts}
\usepackage{algorithmic}
\usepackage{graphicx}
\usepackage{textcomp}
\usepackage{xcolor}
\usepackage{booktabs} 
\usepackage{times}
\usepackage{epsfig}
\usepackage{multirow}
\usepackage{hhline}
\usepackage{tikz}
\usepackage{enumitem}
\usepackage{ulem}
\usepackage{caption}
\usepackage{url}
\usepackage{hyperref}
\usepackage{color, colortbl}
\usepackage[rightcaption]{sidecap}

\def\BibTeX{{\rm B\kern-.05em{\sc i\kern-.025em b}\kern-.08em
    T\kern-.1667em\lower.7ex\hbox{E}\kern-.125emX}}

\definecolor{lightblue}{rgb}{0.8,0.93,1}

\newcommand{\sys}{\textsc{Hashtag}}

\graphicspath{{figures/}} 
\newcolumntype{B}{>{\columncolor{lightblue}}c}

\begin{document}

\title{\sys{}: Hash Signatures for Online Detection of Fault-Injection Attacks on Deep Neural Networks
}

\author{\IEEEauthorblockN{Mojan Javaheripi, Farinaz Koushanfar}
\IEEEauthorblockA{\textit{Department of Electrical and Computer Engineering, UC San Diego} \\
\href{mailto:mojan@ucsd.edu}{\color{blue}mojan@ucsd.edu}, \href{mailto:farinaz@ucsd.edu}{\color{blue}farinaz@ucsd.edu}}
}

\maketitle

\begin{abstract}
We propose \sys{}, the first framework that enables high-accuracy detection of fault-injection attacks on Deep Neural Networks (DNNs)  with provable bounds on detection performance. Recent literature in fault-injection attacks shows the severe DNN accuracy degradation caused by bit flips. In this scenario, the attacker changes a few weight bits during DNN execution by tampering with the program's DRAM memory. 
To detect runtime bit flips, \sys{} extracts a unique signature from the benign DNN prior to deployment. The signature is later used to validate the integrity of the DNN and verify the inference output on the fly. We propose a novel sensitivity analysis scheme that accurately identifies the most vulnerable DNN layers to the fault-injection attack. The DNN signature is then constructed by encoding the underlying weights in the vulnerable layers using a low-collision hash function. When the DNN is deployed, new hashes are extracted from the target layers during inference and compared against the ground-truth signatures. \sys{} incorporates a lightweight methodology that 
ensures a low-overhead and real-time fault detection on embedded platforms. Extensive evaluations with the state-of-the-art bit-flip attack on various DNNs demonstrate the competitive advantage of \sys{} in terms of both attack detection and execution overhead.
\end{abstract}


\section{Introduction}\label{sec:intro}

Deep Neural Networks (DNNs) have enabled a transformative shift
in various applications ranging from natural language processing and computer vision to healthcare and autonomous driving. With the deep integration of autonomous systems in safety-critical tasks, model assurance and decision robustness have gained imminent importance~\cite{javaheripi2020curtail,javaheripi2020cleann}. Although DNNs demonstrate superb accuracy in controlled settings, it has been shown that they are particularly vulnerable to fault-injection attacks. Recent work~\cite{rakin2019bit,hong2019terminal} demonstrates how changing a few bits of the victim DNN's weights can reduce the classification accuracy to below random guess. These malicious bit flips have been realized in DNN accelerators via rowhammer attacks on the DRAM containing the model weights~\cite{yao2020deephammer}. 

In response to bit-flip attacks, prior work suggests adding specific constraints on DNN weights during training such as binarization~\cite{rakin2021ra}, clustering~\cite{he2020defending}, or block reconstruction~\cite{li2020defending}. Adding such constraints increases the number of bit-flips required to deplete the inference accuracy, however, they do not entirely mitigate the threat. Additionally, the proposed constraints often severely affect the underlying DNN's test accuracy. Other work~\cite{li2020deepdyve,liu2020concurrent} propose to use machine learning (ML) based techniques where a simpler model is trained to detect faults in the victim DNN. However, their detection rate and false positive rate are bound by the accuracy of the ML-based detector. 
To ensure DNN robustness, it is crucial to augment autonomous systems with an online fault detection strategy that delivers strict performance guarantees.
To the best of our knowledge, none of the earlier works
provide the needed detection strategy.

We propose \sys{}, a highly accurate real-time fault detection methodology for DNNs deployed in embedded applications. \sys{} is the first method to provide strict statistical bounds on fault detection performance and deliver $0\%$ false positive rate. \sys{} extracts a unique signature from the benign DNN prior to deployment. At runtime, the signature is used to validate the integrity of the DNN and verify the inference output on the fly. We propose to leverage a low-collision hashing scheme, called the Pearson hash, to extract an 8-bit signature from the pertinent weights in each DNN layer. 
Our hash-based signature extraction delivers several benefits: (1)~hash-based integrity check enables accurate fault detection that is robust to false alarms. (2)~The hash algorithm is devised particularly for low-overhead execution on commodity processors. 

There exist an inherent trade-off between fault detection performance and the storage/runtime overhead that is determined by the number of DNN layers used for signature extraction. To balance this trade-off, we propose a novel sensitivity analysis scheme that identifies the most vulnerable layers within the DNN to be used for signature extraction. This, in turn, leads to an extremely lightweight detection methodology that incurs negligible storage and runtime, making it amenable for use in resource-constrained embedded environments. Notably, our sensitivity analysis enables \sys{} to achieve a $100\%$ detection rate using as few as one layer for hash extraction.

Our detection strategy is compatible with the challenging threat model where the attacker has full control over the DRAM to freely select the location and number of bit flips. In addition, the attacker has full knowledge of the underlying detection algorithm, i.e., the hash function. To calibrate \sys{} detection, the user does not require access to any labeled data, fine-tuning, or model training. The user only chooses a secret reordering rule to generate the input for the hash function from the DNN layer weights. Using the reordering rule, the hash signatures can be robustly extracted from the DNN at runtime without the attacker's interference.

We validate the effectiveness of \sys{} by performing extensive
experiments on various DNN architectures and visual datasets. The evaluated DNNs are injected with the state-of-the-art progressive bit-flip attack~\cite{rakin2019bit}. We show that \sys{} achieves a $100\%$ detection rate with $0$ false alarms while incurring $<1.3KB$ storage and $<1\%$ runtime compared to DNN inference on an embedded GPU. Our proposed methodology outperforms prior art across all benchmarks both in terms of attack detection and algorithm execution overhead. Compared to best prior work, \sys{} shows orders of magnitude faster execution and lower storage.

\noindent In summary, the contributions of \sys{} are as follows:
\begin{itemize}
    \item Introducing \sys{}, the first framework for online detection of DNN fault-injection attacks with provable guarantees on performance. 
    \item Constructing a novel signature generation scheme based on Pearson hash which enables low-overhead and highly accurate fault detection.
    \item Providing lower bounds on attack detection rate using a statistical analysis of hash collision.
    \item Devising a sensitivity analysis to identify vulnerable layers within any given DNN architecture. \sys{} automatically finds DNN layers with a high probability for attack and tailors the fault detection to those layers.
\end{itemize}
\section{Background and Prior Work}\label{sec:prelim}
\subsection{Bit-Flip Attack}\label{sec:bfa}
Recent work has developed various fault-injection techniques~\cite{kim2014flipping,tatar2018throwhammer,van2016drammer} that can be utilized to alter bits stored in the DRAM memory. These techniques give rise to the plethora of attacks that take advantage of the bit-flipping tools to induce adversarial behavior in deployed DNNs. Researchers have demonstrated the vulnerability of DNNs to fault-injection attacks that target model parameters. Perhaps the pioneer in this domain is~\cite{liu2017fault} which alters a single parameter throughout the DNN to change the classification result. Follow-up work~\cite{hong2019terminal} analyzes the effect of targeted bit flips induced by the Row hammer attack on DNN accuracy. The authors perform the bit flips in the floating-point representation and show that their injected bitwise errors can lead to $>90\%$ accuracy degradation when applied on certain DNN parameters. 

Current state-of-the-art bit-flip attack~\cite{rakin2019bit} leverages a gradient-based progressive bit search to strategically identify the vulnerable bits in the DNN. Their attack is applied on quantized DNN parameters with the fixed-point representation. 
Other variants of the bit-flip attack exist which leverage a similar adaptive method to find the vulnerable bits but differ in the attack objective: rather than degrading the accuracy on all samples, authors of~\cite{rakin2020tbt,rakin2020t} perform bit flips to misclassify certain input examples as a target class.
In this paper, we direct our focus to the generic untargeted bit-flip attack~\cite{rakin2019bit,yao2020deephammer} as it provides the most general attack objective. We emphasize that \sys{} is applicable to other attack variants as our methodology relies on signature extraction and verification. This, in turn, allows us to detect (adversarial) changes in DNN parameters regardless of the underlying attack objective.

\noindent \textbf{Attack Formulation.} Let us denote by $\{B_l\}_{l=1}^L$ the total bits from the Two's complement representation of per-layer DNN weights where $l$ is the layer index. 
To maximally reduce the DNN accuracy, the attacker iteratively identifies the bit with the highest gradient $\max_{B_l}|\nabla_{B_l}\mathcal{L}|$ in each layer of the DNN. Here, $\mathcal{L}$ denotes the DNN inference loss. 
Once the per-layer most vulnerable bits are detected, the new loss will be measured for each candidate bit-flip. Finally, the bit that results in the maximum loss is selected and flipped. 
The iterative process continues until the DNN accuracy falls below the attacker's desired value. 


\subsection{Existing Defenses}\label{sec:related}

Prior art propose various techniques to increase robustness to fault-injection attacks that occur during DNN training and execution. To thwart training-time attacks, authors of~\cite{trust-CPS21}, propose a trust-based framework as the fault detection mechanism. The performance of this method is strongly dependant on the accuracy of the trust evaluation mechanism~\cite{trust19,trust20}. In this paper, we direct our focus to fault injection attacks applied on the DNN's internal parameters at inference time. 

Several prior defenses against inference-time fault injection attacks suggest adding specific constraints to the model during training. Authors of~\cite{he2020defending} show that adding a piece-wise clustering constraint to the training objective or performing binarized training can improve resiliency. Follow-up work~\cite{li2020defending} proposes to locally reconstruct DNN weights during inference to minimize or defuse the effect of the bitwise error caused by the bit flips. Such methods increase the number of bit flips required to reduce the victim DNN's classification accuracy. However, they do not detect or prevent fault-injection attacks. Additionally, due to the added constraints on the pertinent DNN, these methods reduce the inference accuracy of the victim model. Compared to these methods, \sys{} does not affect the inference accuracy in any way and is able to detect the occurrence of bit flips with $100\%$ accuracy.

Other works suggest adding an ML-based attack detection mechanism. Authors of~\cite{li2020deepdyve} train a smaller, checker network to verify the classification results produced by the original DNN. In case of a mismatch, the task is repeated and the output of the victim DNN is accepted, which results in a low detection rate. Compared to \sys{} lightweight detection method, the checker DNN incurs a higher computational/storage overhead and can itself be subject to fault-injection attacks. Another work~\cite{liu2020concurrent} uses the magnitude of the gradient to find sensitive weights. The authors then train a binary classifier on the sensitive weights to find bit flips. The ML-based detection techniques are bound by the classification accuracy of the underlying detector model and thereby have lower true positive rate and higher false positive rate compared to \sys{}. We provide a probabilistic lower bound on \sys{} detection performance that outperforms prior work.



Most recently, authors of~\cite{li2021radar} employ checksums to detect bitwise errors in weight groups. The detection performance of the proposed methodology relies on the choice of the group size, i.e., the number of weights used to compute each checksum value. To obtain a good trade-off between detection performance and the storage/runtime overhead, the authors suggested using higher group sizes. From a probabilistic point-of-view, checksum on large groups has higher false negative rate compared to our hash-based mechanism. This is because checksum inherently overlooks specific even-numbered bit flips. As shown in our experiments, the best reported results from~\cite{li2021radar} achieve lower detection accuracy compared to \sys{} while requiring higher storage and runtime.
\section{\sys{} Methodology}\label{sec:overview}
\begin{figure}[t]
    \centering
    \includegraphics[width=0.95\columnwidth]{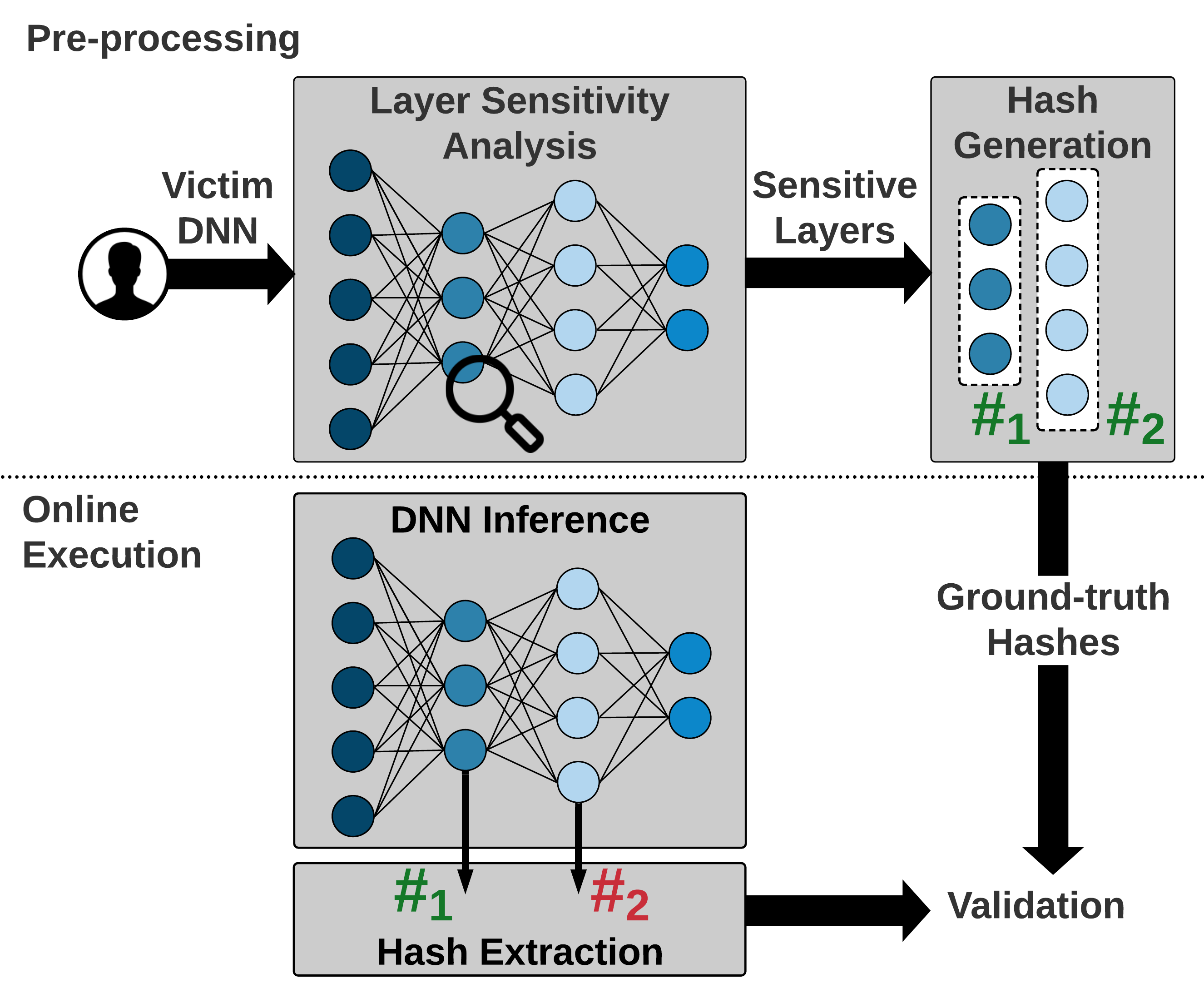}
    \caption{Global flow of \sys{} detection. During the pre-processing phase, we generate a customized signature from a selected subset of DNN layers. During online execution, the signature is used to validate the model's integrity in real-time.}
    \label{fig:global_flow}
\end{figure}

Figure~\ref{fig:global_flow} demonstrates the high-level overview of \sys{} methodology for detecting fault-injection attacks in DNN parameters, i.e., bit flips. The core idea in \sys{} is to generate a compact (ground-truth) signature from the benign DNN. This is done by generating per-layer hashes of DNN parameters prior to model deployment. The signature is then used to verify the integrity of DNN parameters during execution to validate the inference result and mitigate malicious behavior. Our detection methodology incurs minimal computation/storage overhead and is devised based on lightweight solutions to enable efficient and real-time execution in embedded systems. \sys{} comprises two main phases to detect anomalies in DNN parameters: 

\noindent\textbf{Pre-processing Phase.} \sys{} preprocessing is a one-time process in which the detection mechanism is calibrated for the underlying victim DNN. There exist an inherent tradeoff between attack detection performance and the computation/storage requirement for extracting layer signatures;
On the one hand, hashing all layers ensures that the detection mechanism can universally adapt to attacks in any subset of layers. On the other hand, hash computation and storage are linear in the number of layers used for detection. We observe that various DNN layers are not equally targeted by fault-injection attacks. Motivated by this, we devise a 
novel sensitivity analysis scheme that models the vulnerability of DNN layers to bit-flip attacks. The top-k most vulnerable layers, called \textit{checkpoint layers}, are then used to extract the hashes. This, in turn, allows \sys{} to maximize detection performance under any given computation/storage budget.

\noindent\textbf{Online Execution.} This recurring phase is activated when the underlying DNN is queried. During online execution, new hashes are extracted from checkpoint layers in parallel to the DNN inference. The new hashes are then validated against the ground-truth hash values from the pre-processing phase to verify the legitimacy of model parameters. Upon hash mismatch, an alarm flag is raised to notify the user that the system is compromised. The user shall then evict the deployed model and reload the ground-truth weights from the source.

\subsection{Threat Model}\label{sec:threat_model}
In this paper, we direct our focus to fault-injection attacks that target DNN parameters, i.e., the bit-flip attack. In this scenario, the attacker has full knowledge of the victim DNN architecture and its parameters. They further know the physical address of the model parameters and have access to a subset of the data used for training the DNN. The attacker uses the data to progressively identify vulnerable weights and flip their value. This is done by performing a Row Hammer Attack (RHA)~\cite{kim2014flipping} on DRAM locations where the model parameter are stored~\cite{hong2019terminal,yao2020deephammer}. To keep the attack stealthy and reduce the high cost of RHA, we assume the attacker is motivated to minimize the number of flipped bits as is observed in the state-of-the-art attacks~\cite{rakin2019bit,rakin2020tbt}. As such, we do not consider random bit flips since they are shown to be ineffective in reducing DNN accuracy even with a high number of flipped weights~\cite{rakin2019bit,yao2020deephammer}.

We evaluate our detection in the challenging white-box scenario where the attacker knows which layers are used for detection. He is also fully aware of the hash algorithm used for generating the per-layer signatures. However, he does not know the secret hash values and the parameter ordering used for generating the hashes. Following prior work~\cite{li2021radar}, we assume the secret hashes are stored in the secure on-chip SRAM which is not accessible by the attacker. Note that even when SRAM storage is not available, our detection secrets are still immune to RHA. This is due to their low memory footprint (less than 5 KB) that makes them hard to target by RHA as shown in~\cite{hong2019terminal}.
\section{\sys{} Components}\label{sec:methodology}

\subsection{Hash-based Signature Extraction}\label{sec:hashing}

Hash functions generate a constant-length code value which is independent of the size of the corresponding hashed data. This property motivated us to leverage hashing as the underlying mechanism for extracting DNN layer signatures. Among the available hash functions, \sys{} incorporates the Pearson hash~\cite{pearson1990fast} which operates on input streams at Byte granularity. Below we present the Pearson scheme for generating an 8-bit hash value.


\noindent\textbf{Pearson Hash Formulation.} The user generates a \textit{hash table} $T$ which contains a random permutation of integer values in the range [0, 255], i.e., $\mathbb{Z}_{256}$.
For an incoming vector of length $N$ containing Byte values $\{x_i\}_{i=1}^N$, the Pearson hash is defined recursively as follows:
\begin{equation}
    h(x_1, x_2, \dots, x_N)= T(h(x_1, x_2, \dots, x_{N-1})\oplus x_N)
    \label{eq:pearson_hash}
\end{equation}
where $\oplus$ represents the \texttt{XOR} operation. Since $T$ is an arbitrary permutation of values in  $\mathbb{Z}_{256}$, there exists a total of $(256)!$ hash variations for a fixed input stream. The Pearson hash can  be extended to generate hashes longer than 8 bits by repeating the above process several times and concatenating the results. However, as shown in our experiments, the 8-bit Pearson hash accurately detects the state-of-the-art bit-flip attack~\cite{rakin2019bit}.

Our hashing scheme provides several desirable characteristics that makes it particularly amenable for low-overhead detection of fault injection attacks: (1)~The hash computation is well-defined for execution in 8-bit processors and embedded CPUs~\cite{pearson1990fast}. (2)~The hashing scheme is applicable to input streams of varying lengths, thereby providing high customizability for various DNN layer configurations. (3)~Pearson hash accommodates input streams with fixed-point representation which have been target to contemporary bit-flip attacks~\cite{rakin2019bit,rakin2020tbt}. 
Fixed-point parameter values are observed in quantized DNNs that are widely deployed in embedded systems. 

\begin{figure}[h]
    \centering
    \includegraphics[width=0.8\columnwidth]{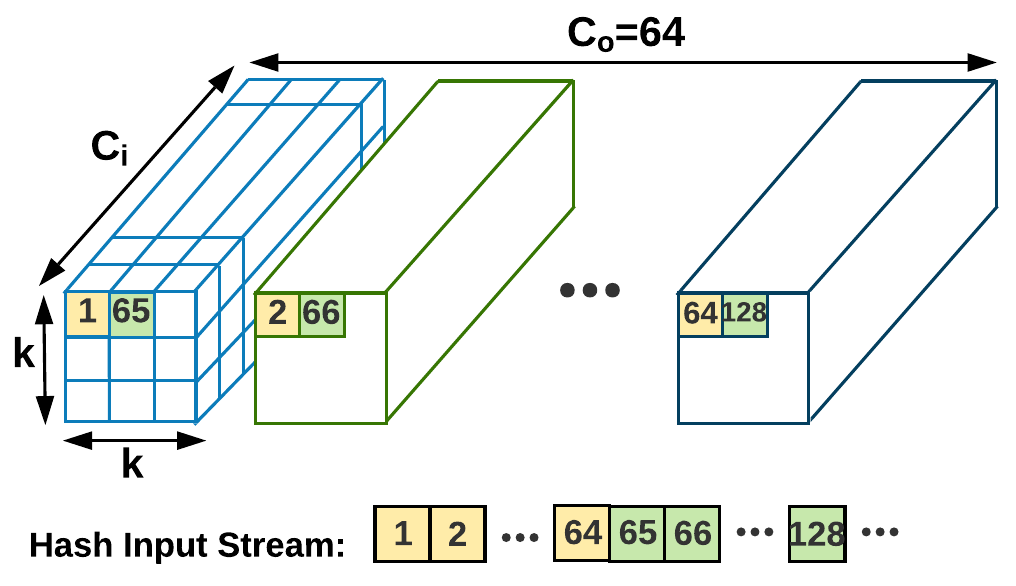}
    \caption{Reordering parameters in an example Convolution layer for generating the hash input stream. The layer parameters are the convolution weight kernels $\in \mathbb{R}^{k\times k\times C_i\times C_o}$ where $k$, $C_i$, $C_o$ denote the kernel size, input channels, and output channels.}
    \label{fig:hash_order}
\end{figure}

\noindent\textbf{Signature Generation.} To extract the ground-truth signature from a benign DNN layer, we first generate a random hash table $T$. The pertinent layer parameters are then fed to  Equation~\eqref{eq:pearson_hash} as the input stream $x_1,x_2,\dots,x_N$ to generate the secret hash of the layer. The hash input stream is generated using a user-defined secret \textit{ordering}. An example of such ordering is shown in Figure~\ref{fig:hash_order}. Here, the hash input stream is constructed by first traversing the layer's weight kernel in the output channel dimension. Ordering adds a zero-cost layer of complexity to \sys{} signature generation which prevents the attacker from reproducing the per-layer secret hashes. Note that the hash input ordering does not affect \sys{} detection performance. The user can easily choose different secret orderings for various layers or change the ordering at any time to reinforce system integrity.


\subsection{Bounds on Detection Performance}
In this section, we provide the worst-case performance bounds on our hash-based detection mechanism. 
Recall from the threat model (Section~\ref{sec:threat_model}) that the attacker is not aware of the secret ordering used to generate the hashes from layer parameters. As such, even if the attacker gains full access to the Pearson hash tables, they will not be able to reproduce the ground-truth hash values. The attacker, therefore, performs the bit-flip attack without taking extra measures to preserve the ground-truth hashes. 
In this context, the lower bound on \sys{} detection can be obtained by quantifying the probability of collision in our hashes. Collision occurs when multiple input streams are mapped to the same output hash.
We analyze hash collision in two separate scenarios where the attacker alters 1)~one or 2)~more than one element of the parameter tensor in the target layer.

\subsubsection{\textbf{Single-element Alteration}} 
When the attacker alters only one element in the weight block where the hash is computed, the user can detect the hash mismatch with $100\%$ accuracy. This is due to an intrinsic collision property for the Pearson hash: for two input streams with exactly one value difference, the probability of collision is zero when the streams are Pearson hashed. 

Let us denote the altered weight value by $\tilde x_m$. The Pearson hash operation for the first $m$ bits can be written as:
\begin{equation}
    h_m = T(h_{m-1}\oplus \tilde x_m)
\end{equation}
where $h_i$ is the short notation for $h(x_1, x_2, \dots, x_i)$. Since the first $m-1$ bits are unaltered, the value of $h_{m-1}$ remains constant. By changing $x_m$, the hash value $h_m$ changes due to the bijective property of the hash table $T$. Since the remaining weight elements $x_i|_{i=m+1}^N$ are unaltered, the new hash $h_m$ propagates through the rest of the input chain, resulting in a different final hash $h_N$ compared to the original weight block. 

\vspace{0.2cm}
\subsubsection{\textbf{Multi-element Alteration}} 
In cases where the attacker changes more than one weight value in the hash block, a possibility arises that the hash mismatch caused by the earlier perturbed elements is later corrected by a subsequent perturbed weight element such that the overall hash value $h_N$ remains unchanged. Without loss of generality let us assume only two elements are altered: $\tilde x_m$ and $\tilde x_n$ ($m<n$). As shown previously, changing the $m^{th}$ element, results in a new hash value that propagates through the input chain until the next changed element. Let us denote the hash value of the first $n-1$ elements in the original and altered weight blocks by $h_{n-1}$ and $\tilde h_{n-1}$, respectively. To ensure the final hash value of the block remains the same, the new value of the $n^{th}$ element $\tilde x_n$ needs to satisfy the following equation:
\begin{equation}
    h_{n-1}\oplus x_n = \tilde h_{n-1}\oplus \tilde x_n
\end{equation}
The above equation limits the number of allowed values for $\tilde x_n$ to only one. As such, the overall probability of obtaining the same hash after altering the bits in two elements is $\frac{1}{256}\sim 0.004$. This probability quantifies the chance of collision occurring in our hashing scheme and remains the same for any arbitrary number of elements altered bigger than one. As such, our (worst-case) lower bound on hash mismatch detection for the DNN is $\big (\frac{1}{256}\big )^{l_a}$. Here, $l_a$ denotes the number of attacked layers where more than one weight element is flipped by the attacker. 

\begin{figure}[h]
    \centering
    \includegraphics[width=0.9\columnwidth]{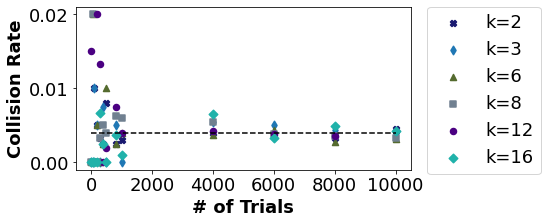}
    \caption{Collision rate versus number of trial runs for hashing an input stream of length $1000$. Each trial randomly changes a subset $k\in[2,3,6,8,12,16]$ of message elements.}
    \label{fig:bounds}
\end{figure}

We empirically evaluate our developed bound by performing multiple runs of hash extraction on an arbitrary input stream of length $1000$. We randomly change a subset of $k$ values within the input and measure the collision rate. As seen in Figure~\ref{fig:bounds}, by increasing the number of experiments, the collision probability asymptotically reaches $0.004$ in all settings, which is compatible with the bound from our statistical analysis.

\subsection{Per-layer Sensitivity Analysis}\label{sec:sensitivity}
State-of-the-art fault injection attacks leverage various techniques to identify weight values that most affect the accuracy if altered. By targeting the attack towards such vulnerable weights, the attacker requires very few bit flips to degrade the accuracy of the victim DNN below random guess. 
Motivated by this, 
we devise a sensitivity analysis that accurately finds the subset of layers inside the victim DNN that are most prone to fault injection. Our sensitivity formulation is inspired by prior work in DNN pruning~\cite{molchanov2019importance}. Specifically, we utilize Taylor expansion to model the effect of per-layer weight change on DNN accuracy as an effective measure of sensitivity. 

Linear layers in DNNs comprise two key parameters, namely the weight and bias: $(W,b)$. Let us represent the entire parameter set for a given DNN with $L$ layers by ${P}=\{(W,b)_1,(W,b)_2,\dots(W,b)_L\}$ where the subscript denotes the layer index. 
Training the DNN is equivalent to minimizing a loss function $\mathcal{L}(D, {P})$ over a corpus of data ${D}={(x_1,y_1),\dots,(x_d,y_d)}$ where $x$ and $y$ correspond to input examples and their labels, respectively. 
To degrade a pretrained DNN's accuracy, the attacker's goal is to maximize the loss over the given dataset. Let us denote by ${P}$ and ${\tilde P}$, the parameters of the DNN before and after the attack. We model the attack objective as:
\begin{equation}
    \max_{{\tilde P}} (\mathcal{L}(D,{P})-\mathcal{L}(D,{\tilde P}))^2 
\end{equation}
We, therefore, quantify the sensitivity of each DNN parameter by the increase in loss value caused by changing it. Bit-flip attacks often alter the sign as it causes the most dramatic change in the value of the underlying parameter, thereby greatly influencing the accuracy~\cite{rakin2019bit}. As such, we model parameter sensitivity by altering the sign $\tilde p = -p$ and measuring the effect on loss. Here the lower case $p$ represents individual weight/bias elements in the DNN. The sensitivity $S(\cdot)$ for the $n^{th}$ parameter $p_n$ can thus be measured as:
\begin{equation}
    S(p_n) = (\mathcal{L}(D,{P})-\mathcal{L}(D,{\tilde P}|_{\tilde p_n=-p_n}))^2 
    \label{eq:sensitivity}
\end{equation}

Since individual computation of \eqref{eq:sensitivity} for each weight element inside the DNN is computationally prohibitive, we leverage Taylor expansion to estimate $S(\cdot)$. For a given function $f(x)$, the first-order approximation using Taylor polynomials at point $x=a$ is given by:
\begin{equation}
    f(x) \approx f(a) + (x-a) \times \frac{\partial f}{\partial x}\biggr\rvert_{x=a}
\end{equation}
By replacing $f$ in the Taylor expansion formula with the loss function $\mathcal{L}$, we can rewrite \eqref{eq:sensitivity} as:
\begin{equation}
    \mathcal{L}(D,{P})-\mathcal{L}(D,{\tilde P}|_{\tilde p_n=-p_n}) \approx 2p_n \times \frac{\partial\mathcal{L}}{\partial p_n}
\end{equation}
We thus measure the sensitivity of parameter $p_n$ as:
\begin{equation}
    S(p_n) \propto (p_n \times \frac{\partial\mathcal{L}}{\partial p_n})^2
    \label{eq:taylor}
\end{equation}

Note that the formula shown in \eqref{eq:taylor} can be easily computed using a simple backward pass through the network to compute the first-order gradients. 
Once the sensitivity is obtained for each weight element, we define the sensitivity of each layer as the average over top-5 sensitivity values of its enclosing elements. We empirically explain our reason for choosing the top-5 weights by providing an analysis of the bit-flip attack in Section~\ref{sec:attack_analysis}

\section{Experiments}\label{sec:exp}
In the following, we provide a comprehensive evaluation of \sys{} performance along with various analyses and discussions. Section~\ref{sec:exp_setup} encloses details of our benchmarked models and datasets, attack setup and implementation, as well as definitions for the utilized evaluation metrics. Section~\ref{sec:attack_analysis} provides an analysis of the attack profile to clarify various design choices. Finally, in Section~\ref{sec:detection} we report the detection performance of \sys{}, provide comparisons with the best prior art, and analyze the storage and computation requirements of \sys{}.

\subsection{Experimental Setup}\label{sec:exp_setup}
\noindent \textbf{Benchmarks.} We evaluate \sys{} on two image datasets, namely, CIFAR10~\cite{cifar10} and ImageNet~\cite{imagenet}. The datasets contain $10$ and $1000$ classes of RGB images of dimensionality $32\times 32$ and $224\times 224$, respectively. We separate $20$ examples from each class in the training data and create a small held-out \textit{validation} dataset. This validation set is used to perform sensitivity analysis in the pre-processing phase.

Table~\ref{tab:dnns} encloses an overview of the DNN architectures evaluated on each dataset and their baseline test accuracies with 8-bit quantization. We evaluate CIFAR10 on two DNNs, namely, ResNet20~\cite{he2016deep} and VGG11~\cite{simonyan2014very}. For ImageNet, we perform experiments on four DNNs, namely ResNet18~\cite{he2016deep}, ResNet34~\cite{he2016deep}, AlexNet~\cite{krizhevsky2012imagenet}, and MobileNetV2~\cite{sandler2018mobilenetv2}.

\begin{table}[h]
\caption{Overview of the evaluated benchmarks. Here, CONV and FC represent Convolution and Fully-connected layer, respectively. The baseline top-1 accuracy and the average number of bit flips are reported for 8-bit quantized DNNs.}\label{tab:dnns}
\setlength\tabcolsep{4pt}
\resizebox{\columnwidth}{!}{
\begin{tabular}{cllcc}
\hline
Dataset                   
& \multicolumn{1}{c}{Model} 
& \multicolumn{1}{c}{Layers} 
& \begin{tabular}[c]{@{}c@{}} Top-1 \\Acc (\%)\end{tabular} 
& Bit Flips
\\ \hline\hline
\multirow{2}{*}{CIFAR10}  
& VGG11                     
& 8 CONV, 3 FC               
& 89.3                                                     
& 90
\\ \cline{2-5} 
& ResNet20                 
& 19 CONV, 1 FC              
& 91.9                                                     
& 18              
\\ \hline
\multirow{4}{*}{ImageNet} 
& AlexNet                   
& 5 CONV, 3 FC               
& 55.5                                                     
& 25              
\\ \cline{2-5} 
& ResNet18                 
& 20 CONV, 1 FC              
& 68.8                                                     
& 8               
\\ \cline{2-5} 
& ResNet34                 
& 36 CONV, 1FC               
& 72.8                                                     
& 9               
\\ \cline{2-5}  
& MobileNet               
& 52 CONV, 1 FC              
& 70.3                                                     
& 3               
\\ \hline
\end{tabular}}
\end{table}

\noindent\textbf{Attack Configuration.}
We leverage the open-source implementation\footnote{Available at \url{https://github.com/elliothe/BFA}} of the state-of-the-art bit-flip attack~\cite{rakin2019bit} to evaluate our detection. The attack batch size is set to $128$ and $64$ for CIFAR10 and ImageNet benchmarks, respectively. Throughout the experiments, we repeat the attack $50$ times with different initial random seeds for each of our DNN benchmarks and report the average obtained results. Each attack round consists of multiple iterations where one bit is flipped at each step. The iterations conclude once the DNN test accuracy falls below the random guess threshold, i.e., $10\%$ and $0.1\%$ for CIFAR10 and ImageNet, respectively. Table~\ref{tab:dnns} encloses the average number of bit flips required for attacking our benchmarked DNNs in the 8-bit quantized regime.

\noindent\textbf{Metrics.} We leverage two evaluation metrics to quantify the performance of \sys{} detection. Firstly, we define Detection Rate (DR) as the ratio of models under attack which are correctly detected by \sys{}, as formulated in Equation~\eqref{eq:dr}. 
\begin{equation}
    DR = \frac{\textrm{\# of attacked models correctly detected}}{\textrm{Total \# of attack rounds}}
    \label{eq:dr}
\end{equation}


Secondly, we use the False Positive Rate (FPR) as the ratio of benign models mistaken for being malicious, i.e., containing a bit-flip that results in a hash mismatch.


\subsection{Attack Analysis}\label{sec:attack_analysis}
\begin{figure}[h]
    \centering
    \includegraphics[width=0.95\columnwidth]{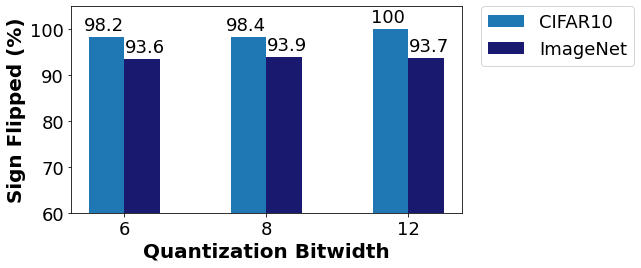}
    \vspace{-0.2cm}
    \caption{Percentage of sign changes occurring during multiple runs of the bit-flip attack. The progressive bit-flip attack~\cite{rakin2019bit} changes the sign of the target parameter with high probability.}
    \label{fig:attack_sign}
\end{figure}

In this section, we perform an ablation study to analyze the characteristics of the bit-flip attack. 
We experiment with two victim DNNs, namely, ResNet20 and ResNet18, trained on the CIFAR10 and ImageNet datasets. The weights in each victim DNN are quantized using a range of bitwidths. The minimum evaluated bitwidth is selected such that the classification accuracy is within $1\%$ and $2\%$ of the floating-point accuracy for CIFAR10 and ImageNet, respectively. For each configuration, we perform $50$ runs of the bit-flip attack with different random seeds to ensure we capture the variances in the outcome. We summarize our findings below:

\noindent \textbf{Sign Change.} Figure~\ref{fig:attack_sign} demonstrates the percentage of bit flips resulting in a sign change across various attack configurations. The consistent pattern among all experiments indicates that the attack significantly favors changing the sign of the target parameter. This is intuitive as flipping the sign of the underlying weight parameter can induce a dramatic change in the output of the layer. Commensurate with this finding, \sys{} sensitivity analysis models the effect of attack as a change in the underlying parameter's sign (See Equation~\eqref{eq:sensitivity}).

\begin{figure}[h]
    \centering
    \includegraphics[width=0.9\columnwidth]{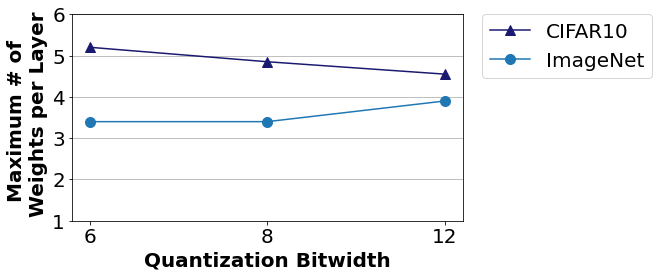}
    \caption{Maximum per-layer attack concentration, averaged across multiple runs of the bit-flip attack. The progressive bit-flip attack~\cite{rakin2019bit} on average targets the weights in the same layer no more than $\sim 5$ times.}
    \label{fig:attack_perlayer}
\end{figure}

\noindent\textbf{Attack Concentration.} To investigate the per-layer attack concentration, we count the number of times each layer is targeted during one execution of the attack. Figure~\ref{fig:attack_perlayer} shows the maximum number of bit flips occurring per layer, averaged across different attack runs. We observe that while the attack could target different or same weights within a certain layer, on average, the same layer is not targeted more than $\sim 5$ times. As such, in our sensitivity analysis, we quantify the sensitivity of each layer as the average over its top-5 sensitive weights.

\subsection{\sys{} Performance}\label{sec:detection}

\subsubsection{\textbf{Sensitivity Analysis}}
\begin{figure}[t]
    \centering
    \includegraphics[width=0.95\columnwidth]{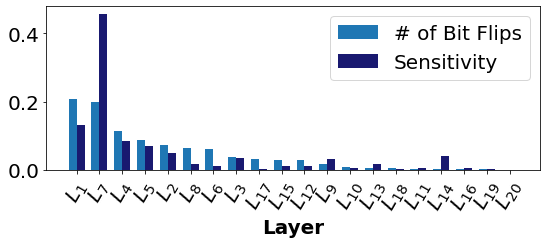}
    \caption{Per-layer sensitivity scores assigned by \sys{} versus the number of per-layer bit-flips. All values are normalized and sum to $1$. Results are gathered across 50 runs of the bit-flip attack on the ResNet20 DNN trained with CIFAR10 dataset.}
    \label{fig:sensitivity}
\end{figure}

In this section, we showcase the stand-alone performance of \sys{} sensitivity analysis. We benchmark the ResNet20 model on CIFAR10 to evaluate the effectiveness of our proposed method in finding the vulnerable layers within a DNN.  
Figure~\ref{fig:sensitivity} demonstrates the sensitivity score assigned to each layer of the model versus the number of per-layer bit flips occurring across $50$ runs of the attack. All values are normalized by the total summation. As seen, there exists a correlation between the sensitivity score and the number of times the pertinent layer has been subject to attack; most attacks occur in layers $1$, $7$ which are also the most sensitive layers found by \sys{}. Below, we provide a thorough evaluation of end-to-end \sys{} execution.

\subsubsection{\textbf{Detection Performance}}
\begin{figure}[b]
    \centering
    \includegraphics[width=0.75\columnwidth]{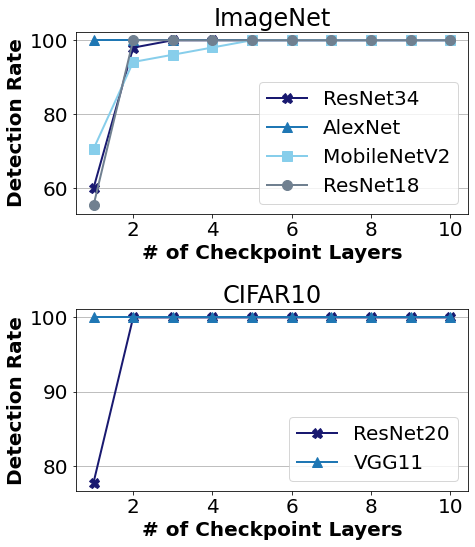}
    \caption{\sys{} detection rate versus the number of checkpoint layers used for signature extraction.}
    \label{fig:detection_rate}
\end{figure}

We leverage our sensitivity analysis to rank DNN layers in the order of their attack vulnerability. 
The top-k most sensitive layers are then selected as checkpoints to extract hashes during the pre-processing and online phases. 
During online execution, if there exists \textit{at least one} hash mismatch with the ground-truth signature among DNN layers, \sys{} marks the model as malicious. 
Figure~\ref{fig:detection_rate} demonstrates the detection performance of \sys{} versus the number of checkpoint layers for various DNN benchmarks. For this experiment, all evaluated models are quantized with 8-bit parameters. 

\sys{} achieves a $100\%$ attack detection rate with very few checkpoints. For the CIFAR10 benchmarks, \sys{} detects faulty DNNs with only $1$ and $2$ checkpoints on the VGG11 and ResNet20 architectures, respectively. For ImageNet, \sys{} achieves a perfect detection rate on AlexNet with only $1$ checkpoint. On the more complex architectures ResNet18 and ResNet34, \sys{} achieves $100\%$ detection with only $2$ and $3$ checkpoints. For the most complex benchmark, i.e., MobileNetV2 with 53 convolution and fully-connected layers, \sys{} achieves $96.2\%$ detection rate with $3$ checkpoints and reaches perfect accuracy with $5$.

The results demonstrate \sys{}'s ability to correctly find the most vulnerable DNN layers and detect fault-injections using hash signatures. Note that \sys{} has a False Positive Rate of {$\mathbf{0.0\%}$}, i.e., it never mistakes benign layers for attacked ones. This is due to the fact that the hash value is constant as long as the underlying layer parameters remain intact, i.e., in the absence of bit flips.

\begin{figure}[h]
    \centering
    \includegraphics[width=0.95\columnwidth]{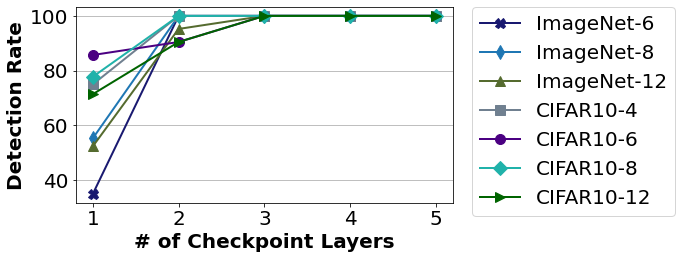}
    \caption{Effect of victim DNN's bitwidth on \sys{} detection rate. The legend presents the utilized datasets along with the underlying bitwidths. ImageNet and CIFAR10 evaluations are performed with ResNet18 and ResNet20, respectively.}
    \label{fig:dr_versus_bits}
\end{figure}

\noindent\textbf{Effect of Bitwidth.} We benchmark ResNet20 and ResNet18 and sweep the quantization bitwidth of the victim DNN. Figure~\ref{fig:dr_versus_bits} demonstrates the effect of DNN bitwidth on \sys{} detection rate. While the bitwidth can affect the detection rate with only one checkpoint, it can be observed that \sys{} becomes agnostic to the underlying bitwidth with more than $2$ checkpoints. For $>2$ checkpoints, \sys{} consistently achieves a detection rate of $100\%$. This property allows \sys{} to be globally applicable to various DNN configurations employed in embedded applications.

\subsubsection{\textbf{Comparison with prior work}}

We compare \sys{} with the best prior work, i.e., WED~\cite{liu2020concurrent} and RADAR~\cite{li2021radar} in terms of detection performance and overhead. We baseline the best reported results in the original papers, i.e., the WED(2) configuration from~\cite{liu2020concurrent}, and $G=8$ and $G=512$ with interleaving for ResNet20 and ResNet18 from~\cite{li2021radar}. We devise two configurations for \sys{} to enable on-par comparison with each of the prior work as follows. 

Similar to \sys{}, the proposed method in~\cite{liu2020concurrent} checkpoints a subset of DNN layers to detect malicious models. 
Therefore, for best comparison with this work, we evaluate \sys{} with the number of checkpoints set to the minimum value required to obtain $100\%$ detection rate (see Figure~\ref{fig:detection_rate}). We call this configuration \texttt{Cfg-1}. 
The method in~\cite{li2021radar}, however, checkpoints all layers within the DNN and reports the performance as the total number of detected bit flips. Therefore, to compare with this work, we devise \texttt{Cfg-2}, where the number of checkpoints is selected such that all bit flips are detected. For \texttt{Cfg-2}, we set the number of checkpoints to $7$ and $8$ for ResNet18 and ResNet20, respectively.

The comparison results are summarized in Table~\ref{tab:prior_art}.
As seen, \sys{} provides state-of-the-art detection performance at a fraction of the storage/computation cost compared to best prior works. Compared to WED~\cite{liu2020concurrent}, \sys{} significantly reduces the false-positive rate and achieves $100\%$ detection rate with $FPR=0.0\%$. Additionally, \sys{} incurs $20-400\times$ lower storage footprint. 
Compared to RADAR~\cite{li2021radar}, \sys{} detects all bit flips within the model with $100\%$ accuracy while incurring $3-4\times$ lower storage cost. We further compare \sys{} runtime with RADAR~\cite{li2021radar}. We measure our runtime on an ARM Cortex-A57 embedded CPU. For a fair comparison, we report the normalized runtimes, i.e., relative to the inference time of the victim DNN on the target hardware. As seen, \sys{} achieves $175-183\times$ faster runtime compared to~\cite{li2021radar}.

We would like to emphasize that unlike~\cite{li2021radar}, \sys{} detection does not rely on the number of detected bit flips. Therefore, the setup in \texttt{Cfg-2} is purely for comparison purposes. The most representative metric for evaluating \sys{} is the detection rate corresponding to \texttt{Cfg-1}, as explained in Section~\ref{sec:exp_setup}, Equation~\eqref{eq:dr}.


\begin{table}[h]
\caption{\sys{} comparison with best prior works WED~\cite{liu2020concurrent} and RADAR~\cite{li2021radar}. Runtime numbers are measured on an ARM CPU and normalized by the inference time of the victim DNN.}\label{tab:prior_art}
\setlength\tabcolsep{1.5pt}
\resizebox{\columnwidth}{!}{
\begin{tabular}{llcccc}
\hline
\multirow{2}{*}{Benchmark}                                      & \multicolumn{1}{c}{\multirow{2}{*}{Work}} 
& \multirow{2}{*}{\begin{tabular}[c]{@{}c@{}}Detection\\ (\%)\end{tabular}} 
& \multirow{2}{*}{\begin{tabular}[c]{@{}c@{}}FPR \\ (\%)\end{tabular}} 
& \multicolumn{2}{c}{Detection Overhead}          
\\ \cline{5-6} 
& \multicolumn{1}{c}{}                      
&                                                              &                                                              & Storage (KB) 
& \multicolumn{1}{l}{Runtime (\%)} 
\\ \hline\hline
\multirow{4}{*}{ResNet20}
& WED                                 
& 96                                                           & 12                                                           & 47           
& N/A  \\
& RADAR                                    
& 97.5                                                         & 0                                                            & 8.2          
& 5.27 \\ \cline{2-6} 
& \texttt{Cfg-1}
& \textbf{100}
& \textbf{0}
& \textbf{0.5}          
& \textbf{0.01}  \\
& \texttt{Cfg-2}
& \textbf{100}
& \textbf{0}
& \textbf{2.1}          
& \textbf{0.03} \\ \hline
\multirow{2}{*}{ResNet18}
& RADAR                                     
& 96.2                                                         & 0                                                            & 5.6          
& 1.83  \\ \cline{2-6} 
& \texttt{Cfg-2}                                 
& \textbf{100}                                                 & \textbf{0}                                                   & \textbf{1.8}          
& \textbf{0.01} \\ \hline
\multirow{2}{*}{ResNet34}
& WED
& 100                                                          & 4                                                            & 302          
& N/A   \\ \cline{2-6} 
& \texttt{Cfg-1}
& \textbf{100}                                                 & \textbf{0}
& \textbf{0.8}          
& $\mathbf{<0.01}$   \\ \hline
\multirow{2}{*}{MobileNet} 
& WED
& 100                                                          & 6                                                            & 26           
& N/A   \\ \cline{2-6} 
& \texttt{Cfg-1} 
& \textbf{100}
& \textbf{0}
& \textbf{1.3}          
&  $\mathbf{<0.01}$ \\ \hline
\end{tabular}}
\end{table}


\subsubsection{\textbf{Storage and Computation Overhead}} Below we provide a more detailed analysis of the storage and runtime specifications of \sys{} detection.
\sys{} storage and computation are linear in the number of checkpoint layers: we compute and store an 8-bit secret hash per checkpoint layer. In addition, the per-layer Pearson  Pearson hash tables each incur a storage cost of $256B$. The Pearson hash tables can be reused among layers, however, here we report the maximum required storage, i.e., when utilizing a unique hash table per checkpoint layer. For $l$ checkpoint layers, the storage overhead of \sys{} is, therefore, $\mathcal{O}(257\times l)B$.

We measure \sys{} runtime on the Jetson TX2 embedded board with an ARM Cortex-A57 CPU and an NVIDIA Pascal GPU. We develop and optimize the 8-bit Pearson hash in C, which is then invoked during DNN execution to detect bit flips. As a baseline, we report the inference time of the victim DNN running on GPU and CPU. The victim DNN is implemented and executed via PyTorch deep learning library. 
Table~\ref{tab:storage_overhead} encloses the runtime and storage of \sys{} across different benchmarks.
The number of checkpoints is set to the minimum value required for a $100\%$ detection rate. As evident from Table~\ref{tab:storage_overhead}, \sys{} delivers perfect detection performance while incurring a negligible storage and computation cost, making it suitable for real-time embedded DNN applications.

\begin{table}[h]
\caption{\sys{} overhead analysis. Here, $\#$ denotes the number of utilized checkpoint layers. }\label{tab:storage_overhead}
\setlength\tabcolsep{3pt}
\resizebox{\columnwidth}{!}{
\begin{tabular}{lccccc}
\hline
\multirow{2}{*}{Benchmark} 
& \multirow{2}{*}{$\#$} 
& \multicolumn{2}{c}{DNN Inference (ms)} 
& \multicolumn{2}{c}{Detection Overhead} \\ \cline{3-6} 
&                                 
& CPU     
& GPU       
& Storage (\%)       
& Time (ms)  \\ \hline\hline
VGG11                      
& 1                               
& 1.5e3 
& 110.7           
& 3e-3               
& 0.1  \\
ResNet20                   
& 2                               
& 661.8           
& 59.4            
& 2e-2               
& 0.1  \\ \hline
AlexNet                    
& 1                               
& 7.9e3 
& 240.7           
& 4e-4               
& 1.3  \\ \cline{2-6} 
ResNet18                   
& 2                               
& 20.9e3 
& 198.5           
& 4e-3               
& 0.7    \\ \cline{2-6} 
ResNet34                   
& 3                               
& 40.8e3 
& 229.7           
& 3e-3               
& 1.8   \\ \cline{2-6} 
MobileNet                
& 5                               
& 2.6e3 
& 182.2           
& 4e-2               
& 0.1   \\ \hline
\end{tabular}}
\end{table}

\section{Conclusion}\label{sec:conclusion}

This paper presents \sys{}, a highly accurate methodology for online detection of fault-injection attacks in DNN parameters. The core idea in \sys{} is to extract a ground-truth signature from the benign model which is then used for verification at inference time. We extract the signatures by encoding DNN layer weights using a low-collision hash function. To minimize detection overhead, we only extract the hashes from a subset of DNN layers where the probability of attack occurrence is high. Towards this goal, \sys{} is equipped with a novel sensitivity analysis that quantifies the vulnerability of DNN layers to bit-flip attacks. 
\sys{} detection strategy provides several benefits: (1)~it delivers $100\%$ detection rate with $0$ false alarms across a variety of benchmarks. (2)~The proposed detection is backed up by provable performance guarantees that provide a lower bound on the detection rate. (3)~\sys{} incurs negligible storage and runtime overhead, enabling accurate fault detection on resource-constrained embedded devices. 
Our lightweight method and realistic
threat model make \sys{} an attractive candidate for practical deployment. Our thorough evaluations corroborate \sys{}’s competitive advantage in terms of attack detection and execution overhead.

\bibliographystyle{IEEEtran}
\bibliography{ref}

\end{document}